\documentclass[letter]{elsart}

\usepackage{graphics}
\usepackage{graphicx}
\usepackage{epsfig}

\usepackage{amssymb}

\begin{document}

\begin{frontmatter}



\title{A linear RFQ ion trap for the Enriched Xenon Observatory}

 
\author[stanford]{B.~Flatt\corauthref{cor}}, 
\corauth[cor]{Corresponding author. Address: Physics Department, 
Stanford University, 382 Via Pueblo Mall, Stanford, 
CA  94305, USA.  Tel: +1-650-723-2946; fax: +1-650-725-6544.} 
\ead{flatt@stanford.edu}  
\author[stanford]{M.~Green},
\author[stanford]{J.~Wodin}, 
\author[stanford]{R.~DeVoe},
\author[stanford]{P.~Fierlinger}, 
\author[stanford]{G.~Gratta}, 
\author[stanford]{F.~LePort}, 
\author[stanford]{M.~Montero D\'{i}ez},
\author[stanford]{R.~Neilson}, 
\author[stanford]{K.~O'Sullivan},
\author[stanford]{A.~Pocar},
\author[stanford]{S.~Waldman\thanksref{sam}},
\thanks[sam]{Now at Caltech, Pasadena CA, USA}
\author[neuchatel]{E.~Baussan}, 
\author[slac]{M.~Breidenbach}, 
\author[slac]{R.~Conley}, 
\author[csu]{W.~Fairbank Jr.}, 
\author[laurentian]{J.~Farine}
\author[slac]{C.~Hall\thanksref{carter}},
\thanks[carter]{Now at University of Maryland, College Park MD, USA}
\author[csu]{K.~Hall}, 
\author[laurentian]{D.~Hallman},
\author[carleton]{C.~Hargrove},
\author[neuchatel]{M.~Hauger},
\author[slac]{J.~Hodgson},
\author[neuchatel]{F.~Juget},
\author[bama]{D.S.~Leonard}, 
\author[slac]{D.~Mackay},
\author[neuchatel]{Y.~Martin},
\author[csu]{B.~Mong},
\author[slac]{A.~Odian}, 
\author[neuchatel]{L.~Ounalli}, 
\author[bama]{A.~Piepke}, 
\author[slac]{C.Y.~Prescott}, 
\author[slac]{P.C.~Rowson}, 
\author[slac]{K.~Skarpaas}, 
\author[neuchatel]{D.~Schenker}, 
\author[carleton]{D.~Sinclair}, 
\author[carleton]{V.~Strickland}, 
\author[laurentian]{C.~Virtue},
\author[neuchatel]{J.-L.~Vuilleuimier},
\author[neuchatel]{J.-M.~Vuilleuimier}, 
\author[slac]{K.~Wamba}, 
\author[neuchatel]{P.~Weber}

\address[stanford]{Physics Department, Stanford University, Stanford CA, USA}
\address[neuchatel]{Institut de Physique, Universit\'e de Neuchatel, Neuchatel, Switzerland}
\address[slac]{Stanford Linear Accelerator Center, Menlo Park CA, USA}
\address[csu]{Physics Department, Colorado State University, Fort Collins CO, USA}
\address[laurentian]{Physics Department, Laurentian University, Sudbury ON, Canada}
\address[carleton]{Physics Department, Carleton Univerisity, Ottawa ON, Canada}
\address[bama]{Dept. of Physics and Astronomy, University of Alabama, Tuscaloosa AL, USA}

\begin{abstract}
The design, construction, and performance of a linear radio-frequency ion trap (RFQ) intended for use in the Enriched Xenon Observatory (EXO) are described.  EXO aims to detect the neutrinoless double-beta decay of $^{136}$Xe to $^{136}$Ba.  To suppress possible backgrounds EXO will complement the measurement of decay energy and, to some extent, topology of candidate events in a Xe filled detector with the identification of the daughter nucleus ($^{136}$Ba).  The ion trap described here is capable of accepting, cooling, and confining individual Ba ions extracted from the site of the candidate double-beta decay event.  A single trapped ion can then be identified, with a large signal-to-noise ratio, via laser spectroscopy.

\end{abstract}

\end{frontmatter}

\section{Introduction}
In the last decade, compelling evidence for flavor mixing in the neutrino sector has clearly shown that neutrinos have finite masses~\cite{oscillations}.  These experiments reveal mass differences between single mass eigenstates, but not their absolute values. The measurement of such masses has become arguably the most important frontier in neutrino physics, with implications in astrophysics, particle physics, and cosmology. $\beta$-decay endpoint spectroscopy measurements provide an increasingly sensitive probe of neutrino mass~\cite{betadecay}. However, a less direct but potentially more sensitive technique is the observation and measurement of the rate of neutrinoless double-beta ($0\nu\beta\beta$) decay~\cite{Elliott-Vogel}.  The discovery of this exotic nuclear decay mode would provide an absolute scale for neutrino masses and establish the existence of two-component Majorana particles~\cite{Majorana}. 

Sensitivity to Majorana neutrino masses in the interesting 10~-~100~meV region is achievable by experiments utilizing a ton-scale $0\nu\beta\beta$ isotope source~\cite{Elliott-Vogel}. This assumes that backgrounds from natural radioactivity, cosmic rays, and the standard-model two-neutrino double-beta ($2\nu\beta\beta$) decay can be sufficiently reduced and understood.    Several proposals exist to perform this daunting task~\cite{Elliott-Vogel}. The Enriched Xenon Observatory (EXO) is designed to identify the atomic species ($^{136}$Ba) produced in the decay process, using high resolution atomic spectroscopy~\cite{EXO_whitepaper}.    This isotope specific ``Ba tagging," working in conjunction with more conventional measurements of decay energy and crude event topology, will potentially provide a clean signature of $0\nu\beta\beta$ decay.

The EXO collaboration is currently pursuing a $0\nu\beta\beta$ detector R\&D program, focusing on a time projection chamber (TPC) filled with xenon enriched to 80\% $^{136}$Xe in liquid (LXe) or gaseous (GXe) phase.  Many of the detector parameters and, in particular, the details of the Ba tagging technique would be different in LXe and GXe.   The ion trap described here is designed to accept, trap, and cool individual Ba ions extracted from a $0\nu\beta\beta$ detector.  While the technique to efficiently transport ions from their production site is still under investigation (and is beyond the scope of this article), the ion trap discussed here is optimized to operate with a LXe detector and a mechanical system to retrieve and inject the ions.  This ion trap is capable of confining ions for extended periods of time ($\sim$~min) to a small volume ($\sim~(500~\mu$m$)^3$), essential for observing single ions via laser spectroscopy with a high signal-to-noise ratio. These properties are required to drastically suppress candidate decays that do not create Ba ions in the TPC, while maintaining a high detection efficiency for Ba-ion-producing events.  In addition, this trap can operate in the presence of some Xe contamination, which is likely in any Ba tagging system coupled to a Xe filled detector.  The ion trap system described here is designed to be appropriately flexible as an R\&D device.  Simplifications and modifications of this system can be adopted for the actual trap to be used in EXO.

\section{Linear RFQ traps and buffer gas cooling}

\begin{figure}[htb!!]
\begin{center}
\includegraphics[width=12cm]{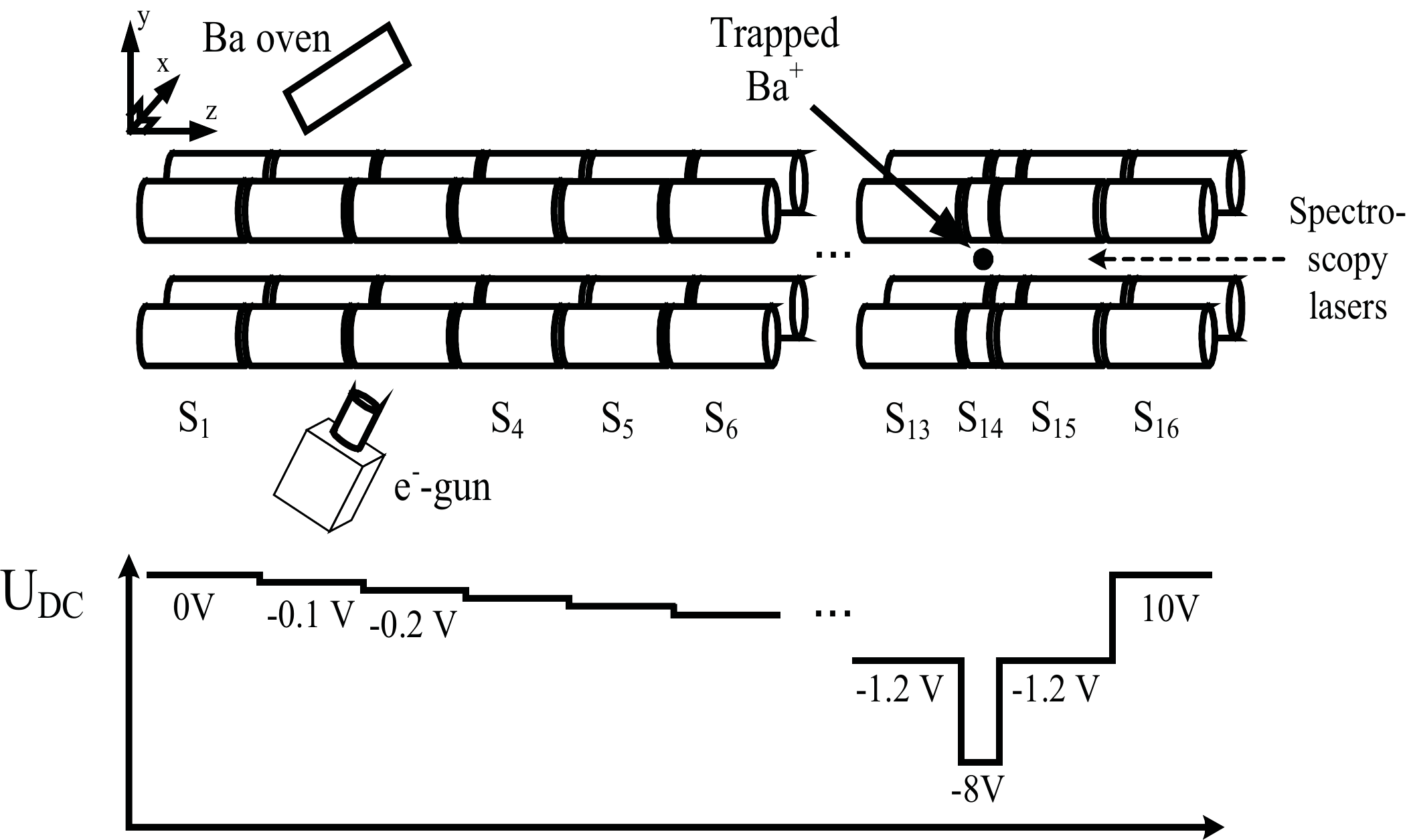}
\vskip 0.1cm
\caption{Schematic of the linear RF trap. Ions are loaded in S$_3$ and stored in S$_{14}$.
The lower part of the figure shows the DC potential distribution.}
\label{linear_trap}
\end{center}
\end{figure}

RF Paul traps confine charged particles in a quadrupole RF field~ \cite{paul}.  Spherical traps have a closed geometry consisting of a ring and two endcap electrodes, providing trapping in three spatial dimensions.  Linear Paul traps generally consist of four parallel cylindrical electrodes, placed symmetrically about a central (longitudinal) axis, as shown in fig.~\ref{linear_trap}.  An RF field applied across diagonally opposing electrodes provides transverse ($x-y$ plane in fig.~\ref{linear_trap}) confinement of the ion.

Appropriately chosen DC voltages, applied to longitudinally ($z$ axis in fig.~\ref{linear_trap}) segmented electrodes, provide longitudinal confinement.  A single group of four symmetrically placed electrodes is referred to as a ``segment".  Electrodes are constructed and positioned such that their radius, $r_e$, is related to the characteristic radial trap size, $r_0$, by
\begin{center}
\begin{equation}
r_e=1.148 r_0
\label{radius}
\end{equation}
\end{center}
where $r_0$ is the distance from the axis of the trap to the innermost edge of an electrode.  This configuration creates the closest approximation to a hyperbolic potential at the trap center for cylindrically shaped electrodes~\cite{Denison}.  The ion's orbit in the transverse plane is described by the Mathieu equation~\cite{Mathieu}. Analysis of the solutions to this equation reveal stability criteria 
for the ion's motion in the trap. The dimensionless Mathieu stability parameters, $a$ and $q$, are defined as
\begin{center}
\begin{eqnarray}
q=2\frac{eU_{RF}}{mr_0^2\omega _{RF}^2}\label{q}\\
a =4\frac{eU_{DC}}{mr_0^2\omega_{RF}^2}\label{a}
\label{traparameters}
\end{eqnarray}
\end{center}
where where $e$ and $m$ are the ion's charge and mass, $\omega_{RF}$ is the angular RF frequency, $U_{RF}$ is the RF voltage (amplitude), and $U_{DC}$ is the DC voltage.  Values of $a$ and $q$ between 0 and 0.91, falling within a region defined by the characteristic numbers of the Mathieu equation, correspond to stable ion orbits~\cite{March}.

Transverse confinement is attributed to a pseudopotential
\begin{center}
\begin{equation}
V_{RF}(r)=\frac{qU_{RF}}{8r_0^2}r^2.
\label{radial} 
\end{equation}
\end{center}
quadratically dependent on the radial distance, $r$, from the longitudinal axis of the trap.  The DC voltages applied to the longitudinally segmented electrodes are chosen to create a trapping potential
\begin{center}
\begin{equation}
V_{DC}(r,z)=\frac{U_{DC}}{z_0^2}\left( z^2-\frac{r^2}{2} \right )
\label{long}
\end{equation}
\end{center}
where $z$ and $z_0$ are the longitudinal coordinate and length of the segment in which the ion is trapped (the ``trapping segment").  The radial dependence of the longitudinal potential well arises from Laplace's equation applied to the interior region of the ion trap.   This radial defocusing reduces the depth of the transverse pseudopotential well, created by the RF field.  The total potential well used to trap ions is the sum of eqns. \ref{radial} and \ref{long}. 

The open geometry of this type of trap allows for an unobstructed view of the trapping region, with large optical angular  acceptance. In addition, the longitudinal electrode segmentation allows for multiple configurations of the longitudinal potentials as required for the injection, trapping, and ejection of a single ion.

In order to confine an energetic ion of mass $m$ injected from outside the trap, a mechanism of energy loss must be provided in order to dissipate the ion's kinetic energy to below both $eV_{RF}$ and $eV_{DC}$. Collisions with a "buffer" gas of mass $m_B$ can provide such an energy-loss mechanism.  The phenomenology of ion-neutral interactions in an RF Paul trap can be divided into three cases.  If $m_B\ll m$, the ion is cooled via a large number of ion-neutral collisions, each exchanging a small amount of energy and momentum. In this  case, the cooling process is adiabatic compared to the period of the ion's motion in the RF field, and the pseudopotential formulation is valid during the cooling process.  If $m_B \gg m$, each collision can add or remove substantial momentum and energy from a trapped ion.  This large instantaneous momentum transfer can alter the ion's trajectory appreciably, which may result in energy transfer from the RF-field to the ion ("RF heating").   Under these conditions, the ion is unstable in the trap, and is rapidly ejected.  In the intermediate regime, $m_B \approx m$, a form of RF heating also occurs and the amount of time that a single ion is trapped depends on trap parameters. 

\section{Simulation of ion cooling and trapping}

The DC and RF voltage amplitudes, the longitudinal dimensions of the individual segments, and the buffer gas pressure and type are optimized using the SIMION~7.0 simulation package\footnote{http://www.sisweb.com/simion.htm} for single ion stability.  Ion-neutral collisions are implemented using a hard-sphere model with a variable radius, depending on the buffer gas and trapped ion species.  This model, applicable in the case of a single atomic ion interacting with a noble buffer gas~\cite{kim}, uses a cross section dependent on the ion's velocity, $v$, and buffer gas polarizability to account for the dipole moment of the neutral atom induced by the ion.  Collisions are implemented by specifying a mean free path $\lambda $, the buffer gas mass $m_B$, and a buffer gas temperature $T_B$.   The probability that a trapped ion collides with a buffer gas atom in a time interval $\Delta t$ is given by

\begin{center}
\begin{equation}
P(\Delta t)=1-e^{-v\Delta t/\lambda}
\end{equation}
\end{center}
Before each time-step of the ion's trajectory, a random number is chosen.  This number is used to decide if a collision occurs during that time-step.  If a collision occurs, the kinematics of the collision are calculated assuming that the velocity distribution of the neutral buffer gas atoms follows Maxwell-Boltzmann statistics with a temperature $T_B$.

The total longitudinal trap length, 604~mm, is chosen as a result of cooling simulations of an ion at various initial kinetic energies, interacting with a range of buffer gases (He, Ar, Kr, and Xe).  The trap is split into 16 segments, to provide sufficient versatility in shaping the longitudinal field for different phases of the R\&D program.   The segments are labeled S$_i$, where $i$ runs from 1 to 16 as shown in fig. \ref{linear_trap}.  The segments are chosen to be 40~mm long, except for a single short 4~mm segment (S$_{14}$, the ``trapping segment"), used to tightly confine the ion longitudinally, optimizing the single ion fluorescence signal-to-noise ratio.

The DC potential profile chosen for most operations is $U_{DC}$ = \{0V, -0.1V, -0.2V, ..., -1.2V, -8.0V, -1.2V, +10V\}, with the minimum of -8~V at S$_{14}$.    Because of a limitation in the number of input parameters of SIMION, this profile is approximated as $U_{DC}$~=~\{+10V, +10V, +10V, -0.4V, -0.5V, ..., -1.3V, -1.2V, -8.0V, -1.2V, +10V\}, in the simulation.    This does not appreciably affect the ion cooling at the potential minimum.    

The values of the trap radius, $r_0$, and electrode diameter, $r_e$ are chosen to optimize the external optical access to a trapped ion, as well as the shape of the RF field.  The electrode radius is $r_e=3$~mm, resulting in $r_0=2.61$~mm (see eqn. \ref{radius}).  The RF frequency is $\omega_{RF}/2\pi=1.2$~MHz, with an amplitude of 150~V.    These parameters correspond to $q=0.52$ and $a=0.05$ (see eqn.~\ref{traparameters}), well within the region of stability of the Mathieu equation.

\begin{figure}[htb!!]
\begin{center}
\includegraphics[width=12cm]{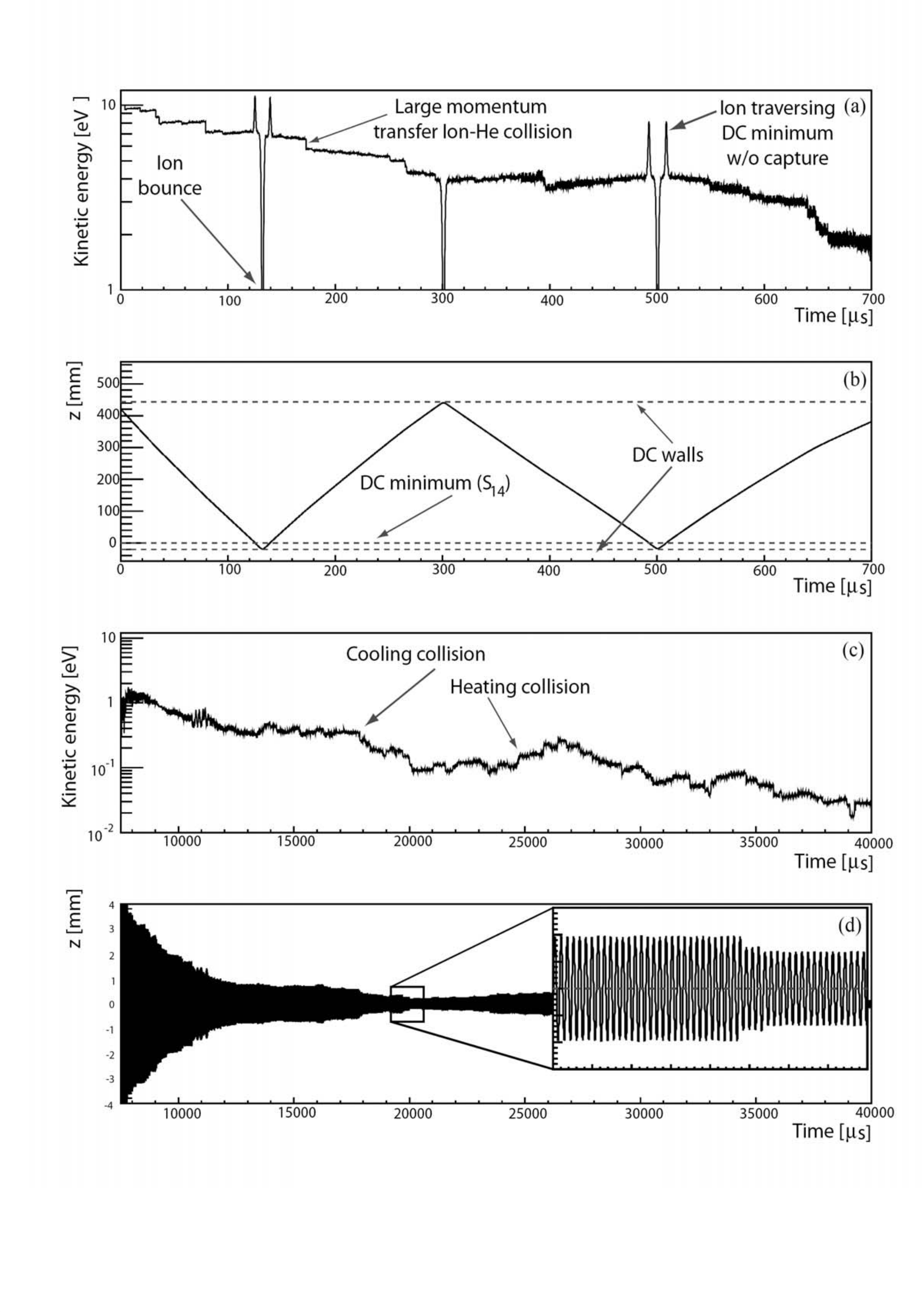}
\vskip 0.1cm
\caption{Simulation of a single $^{136}Ba^+$ in He ($1\cdot 10^{-2}$torr) using SIMION. 
The ion in the simulation was started in the center of segment S$_3$ of the trap.
Panel (a) shows the collisional cooling during the first few hundred 
$\mu $s after the start of the simulation. Panel (b) shows the respective trajectory. 
Panels (c) and (d) show the evolution of the same quantities on a longer time scale.
The ion is confined to the trapping segment and it cools 
further down to the buffer gas temperature.}

\label{simulation_He}
\end{center}
\end{figure}

\begin{figure}[htb!!]
\begin{center}
\includegraphics[width=12cm]{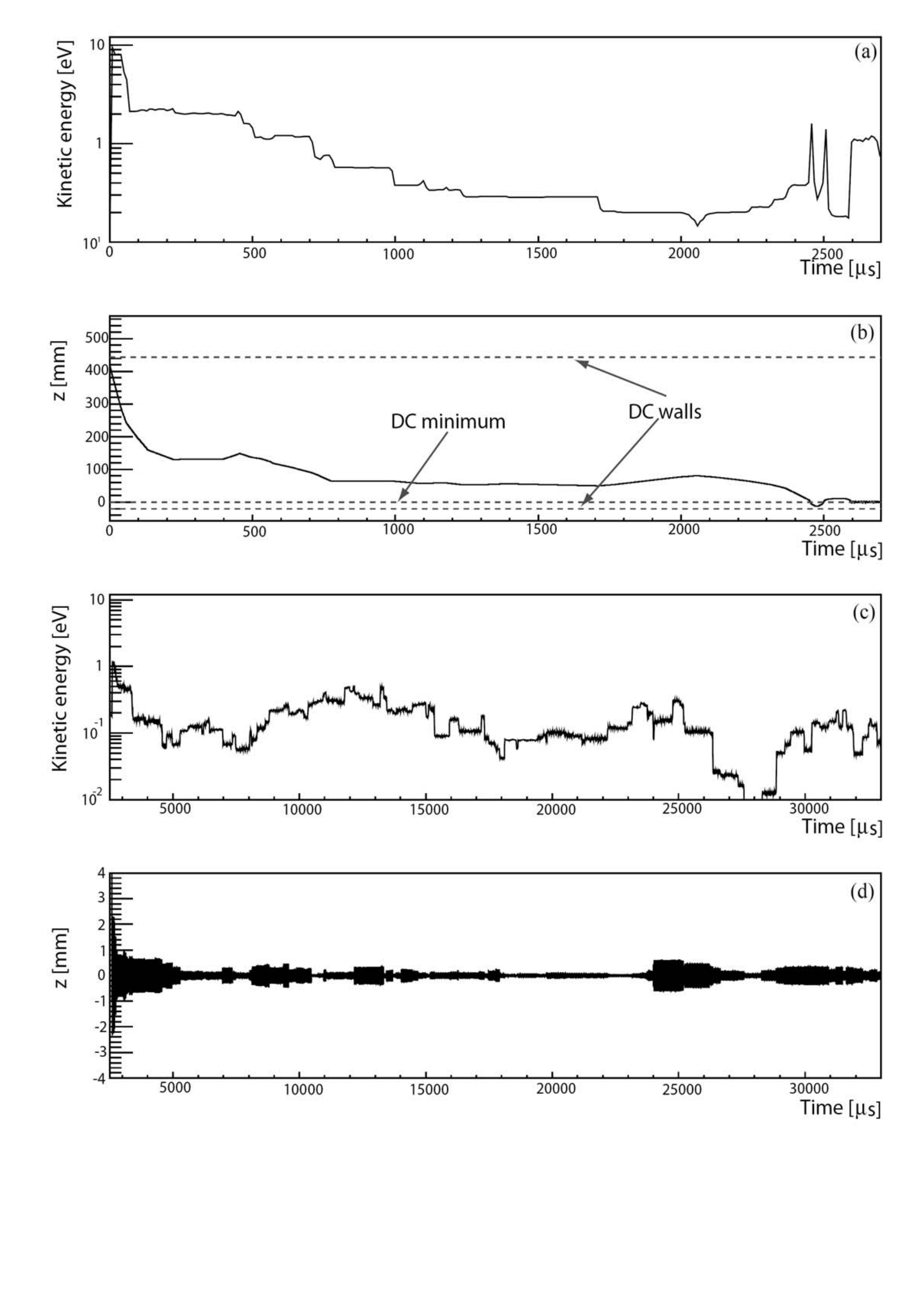}
\vskip 0.1cm
\caption{The same simulation as in fig.\ref{simulation_He} of a single $^{136}$Ba$^+$ in 
Ar at a pressure of $3.7\cdot 10^{-3}$torr. Faster cooling and larger momentum transfer collisions are evident.}

\label{simulation_Ar}
\end{center}
\end{figure}

An example of the simulated cooling process is shown in fig. \ref{simulation_He}.  In this simulation, a single Ba ion is created in S$_3$, with an energy of 10~eV in $1\times10^{-2}$~torr He.  The ion's kinetic energy and $z$-trajectory are plotted during the initial cooling (panels $a$ and $b$), and after the ion is confined in the potential well at S$_{14}$ (panels $c$ and $d$).  During the initial cooling, the ion is reflected back and forth longitudinally in the trap.  On average,  the ion loses energy with each collision with a He atom.  Once the ion is confined to S$_{14}$, the ion continues to cool to the minimum, until it comes into thermal equilibrium with the buffer gas.  The same processes are shown in fig.~\ref{simulation_Ar} in the case of Ar as a buffer gas.  The ion cools much faster in the presence of Ar; however, the frequency and amplitude of RF heating collisions increase as well.  Ar is therefore a more efficient cooling gas for Ba ions, though the higher rate of RF heating likely decreases the ion's stability in the trap.  Whereas SIMION is useful for studying these cooling and heating processes, reliable trajectory simulation is limited to the timescale of a few seconds.  This is due to finite computational resources, as well as error buildup during trajectory integration.  For this reason, single ion storage times longer than a few seconds, relevant for the study of RF heating and ion deconfinement, cannot be simulated.

\section{Trap construction}
\begin{figure}[htb!!]
\begin{center}
\includegraphics[angle=-90,width=12cm]{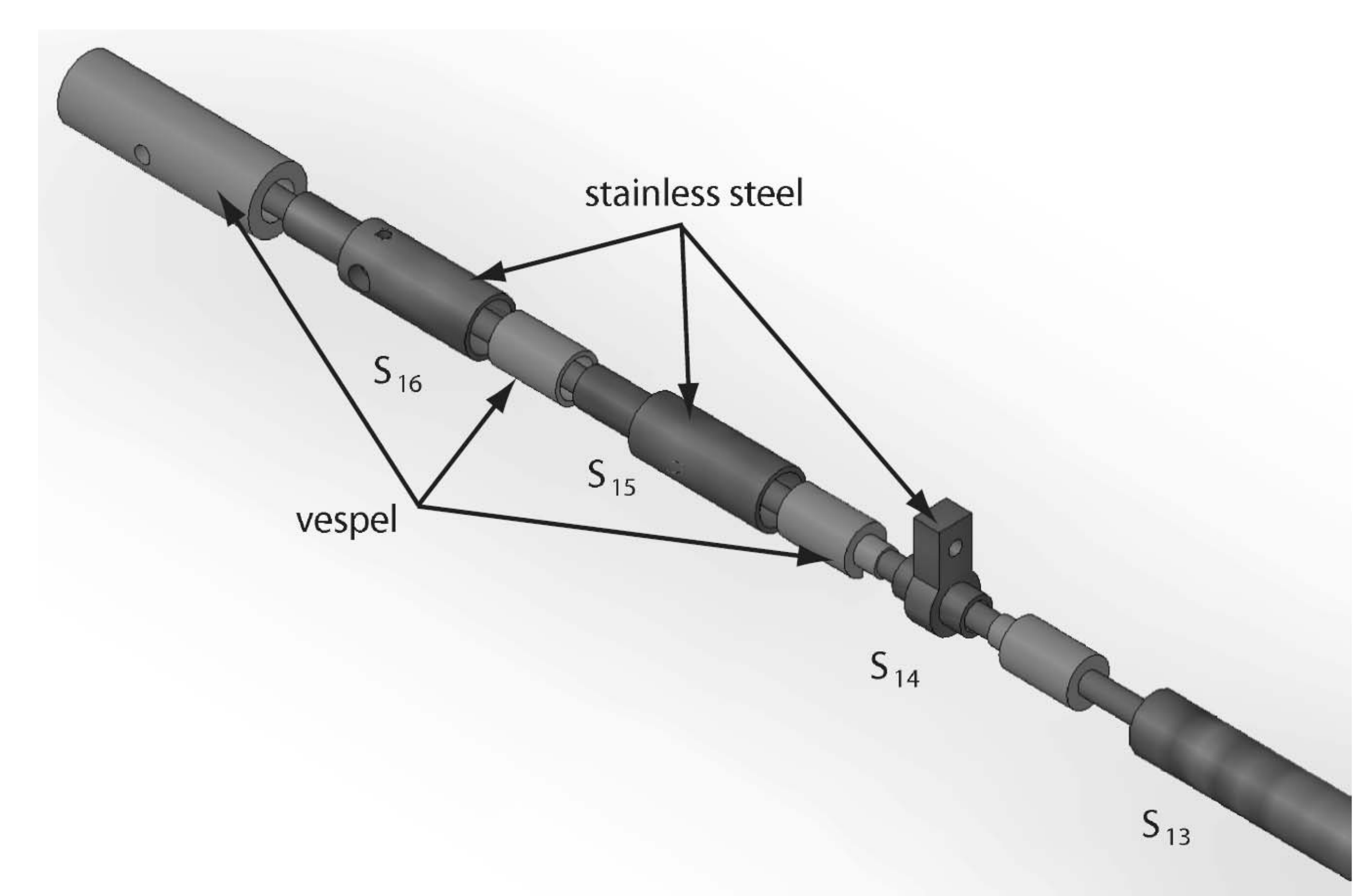}
\vskip 0.1cm
\caption{Exploded view of electrodes 13-16, showing the internal support and electrical insulation.}
\label{electrodes}
\end{center}
\end{figure}

\begin{figure}[htb!!]
\begin{center}
\includegraphics[width=12cm]{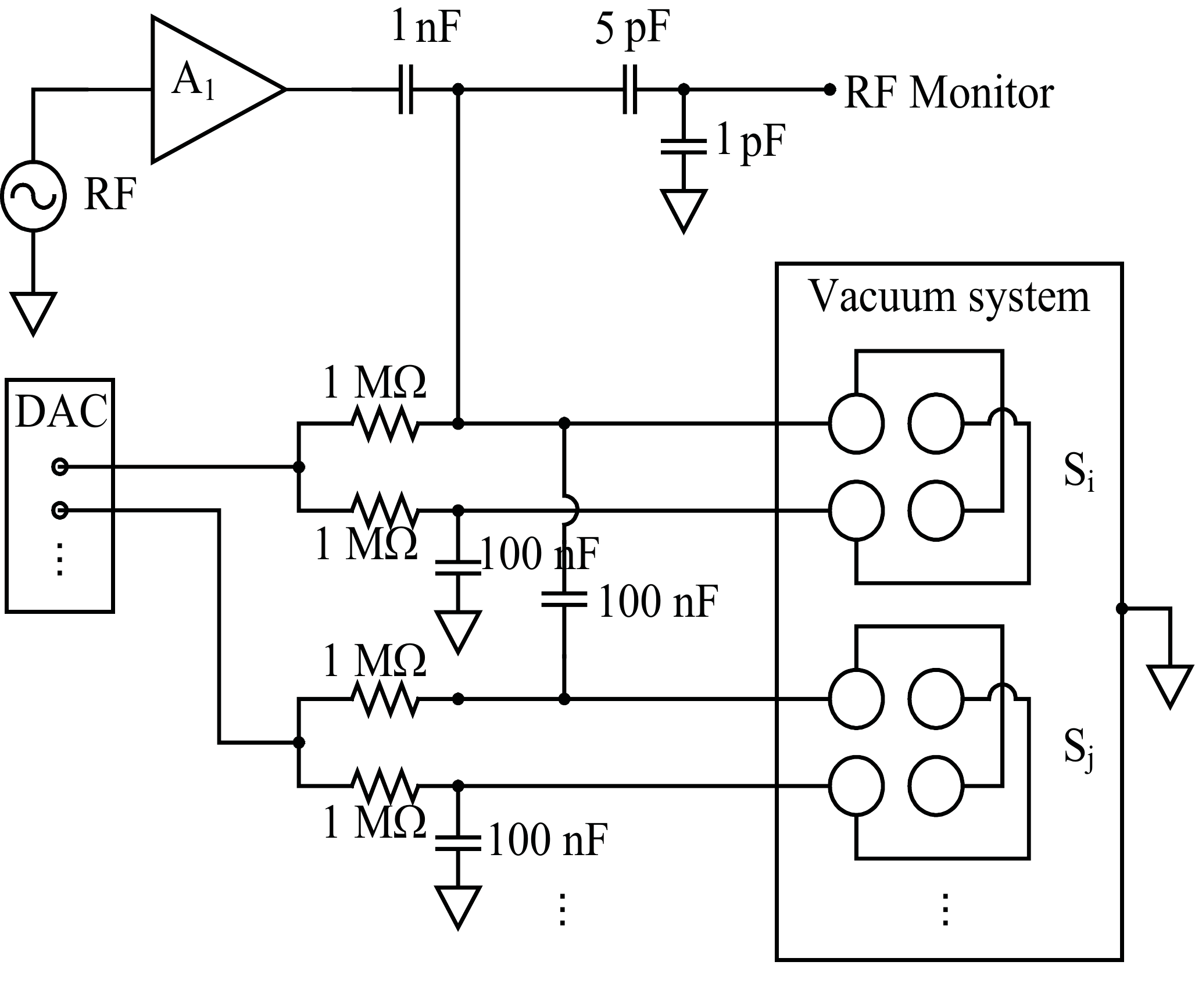}
\vskip 0.1cm
\caption{Electrical schematics of the ion trap. Only two of 16 identical segments are shown.}
\label{RFDC_circuit}
\end{center}
\end{figure}

A single trap segment is made of four stainless steel tubes threaded onto a center stainless steel rod, as shown in an exploded view in fig.~\ref{electrodes}.  Vespel\footnote{Vespel is a trademark of DuPont de Nemours} tube spacers insulate each segment from its neighbors, and from the center rod.  Special care is taken to insure that all vespel parts are recessed behind conductors, in order to avoid any insulator charging that could affect the DC field inside the trap.  Details of the RF and DC feed circuitry for two segments are shown in fig.~\ref{RFDC_circuit}.  A DC voltage is applied to all four electrodes in a segment, using a 16-bit computer controlled DAC.  The RF is applied to one diagonal pair of electrodes in a segment, while the other pair is RF grounded through a capacitor.  The RF signal is supplied by a function generator\footnote{HP 8656B}, which is amplified by a broadband 50~dB amplifier\footnote{ENI A150}, internally back-terminated with 50~$\Omega$.  The system can deliver the RF voltage required without the use of a tuned circuit. Each segment has a capacitance of $\sim$~18~pF, however the total capacitance of the trap is closer to 600~pF due to contributions from cables and vacuum feedthroughs.

\begin{figure}[htb!!]
\begin{center}
\includegraphics[width=13cm]{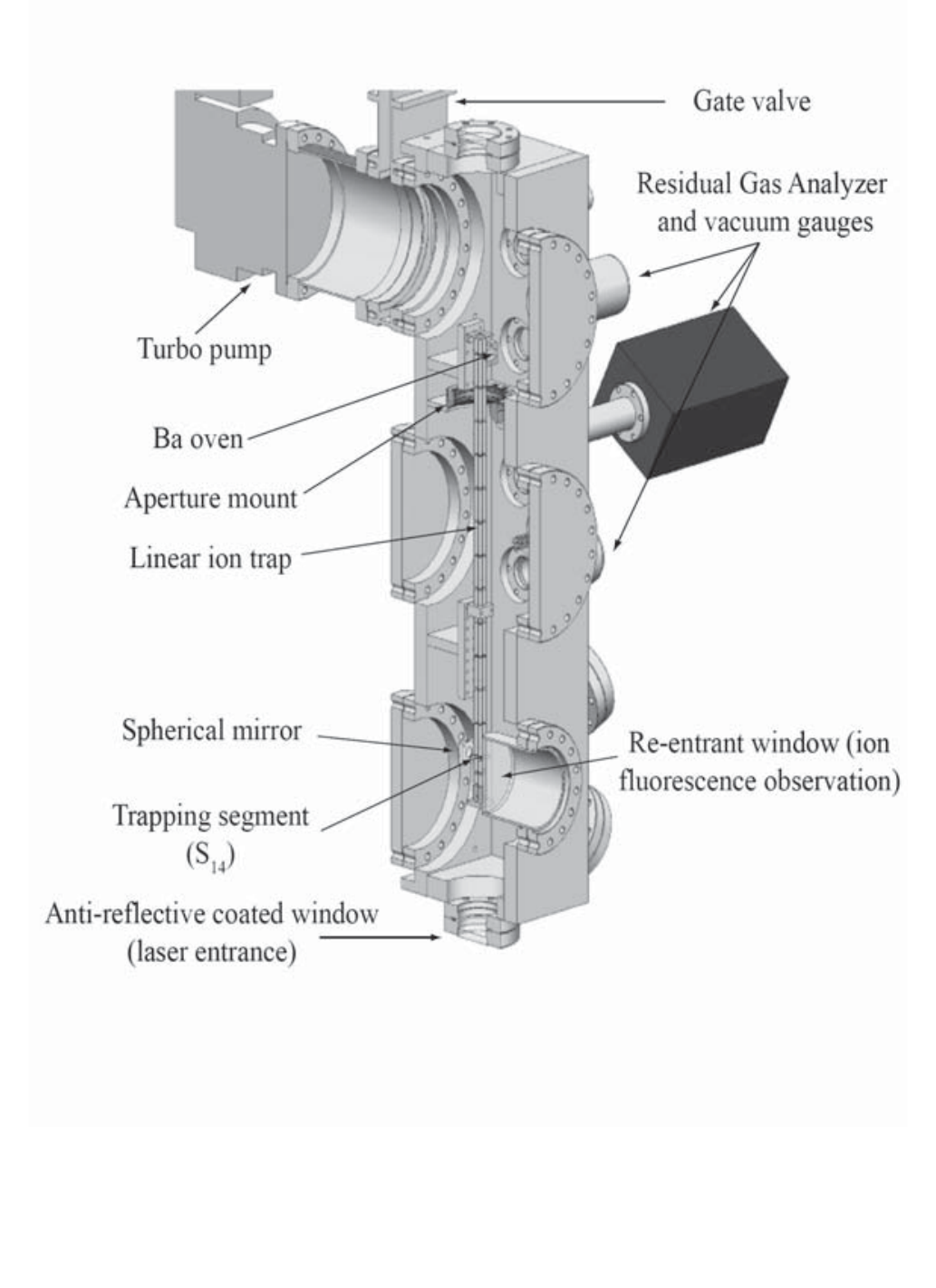}
\vskip 0.1cm
\caption{Cutaway view of the vacuum system with the linear trap mounted inside.}
\label{vac_overview}
\end{center}
\end{figure}

\begin{figure}[htb!!]
\begin{center}
\includegraphics[width=13cm]{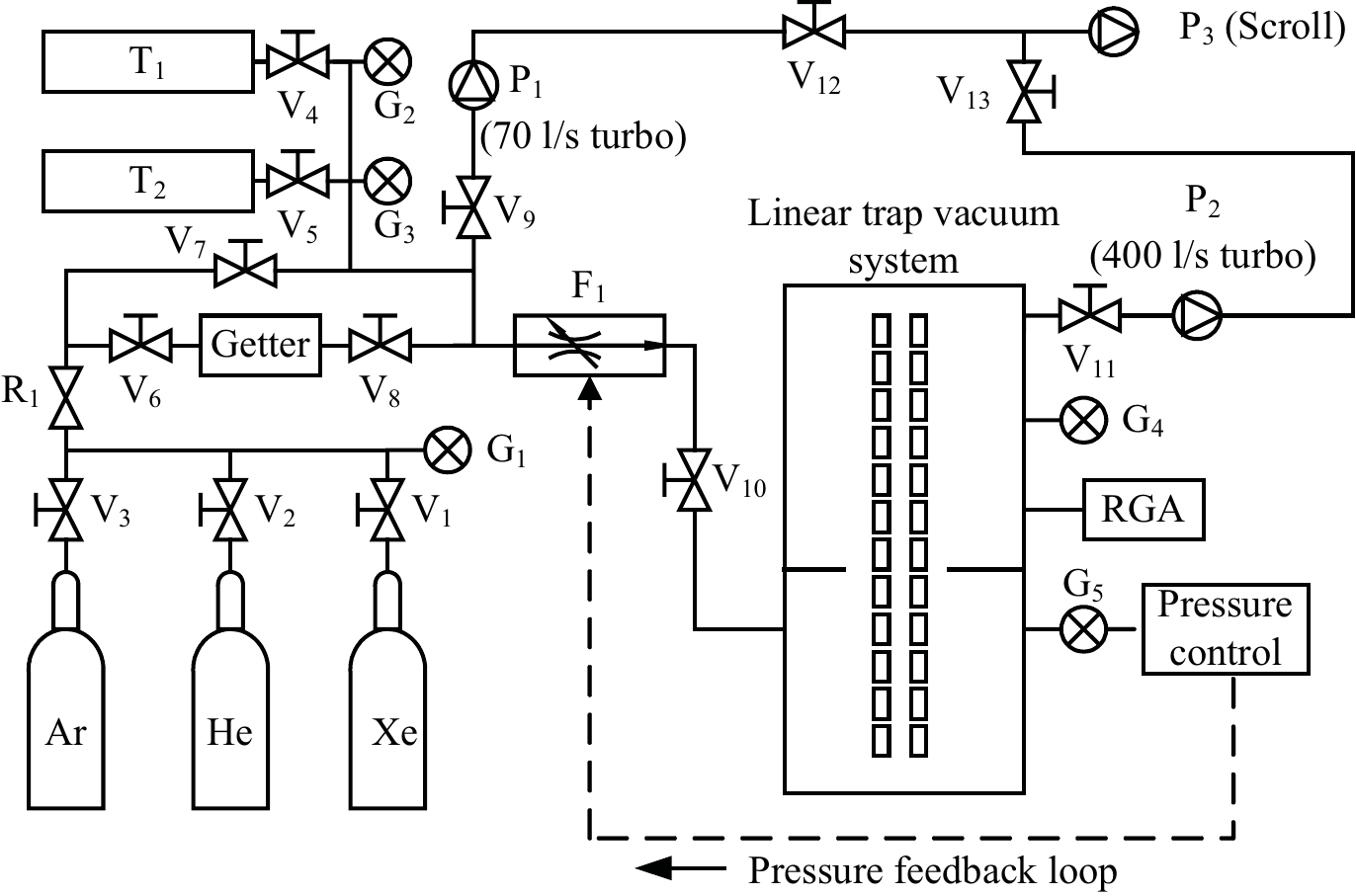}
\vskip 0.1cm
\caption{Schematic view of the gas handling system.}
\label{gashandling}
\end{center}
\end{figure}

The whole trap is housed in a custom-made, electropolished stainless steel UHV tank pumped by a turbomolecular pump\footnote{Pfeiffer 
TMU521P} backed by a dry scroll pump\footnote{BOC Edwards XDS5}(fig.~\ref{vac_overview}).  A septum inside the vacuum tank allows for the 
installation of an aperture, to be used in a differentially pumped scheme (not used for the work described here), to maintain different 
buffer gas pressures in the injection and trapping regions of the system.  The pressure in the tank is read out in the upper (injection) and 
lower (trapping) regions of the vacuum system by vacuum gauges\footnote{Pfeiffer PKR251}. A gas handling manifold, connected to the trap by 
a computer-controlled leak valve\footnote{Pfeiffer EVR116}, allows for the introduction of individual buffer gas species and binary mixtures 
(fig.~\ref{gashandling}).  The leak valve keeps the buffer gas pressure in the vacuum chamber constant by regulating gas flow into the 
chamber, based on the vacuum gauge measurement closest to S$_{14}$.  The turbo pump runs continuously, so that the gas pressure can be 
either increased or decreased at any time.  Using this method, the gas pressure in the vacuum system can be regulated between 
$3\times10^{-9}$ and $1\times10^{-2}$~torr, with a stability of $\leq$~1~\%.  The lower range is the limit of the vacuum gauges, while the 
upper limit is the maximum allowable pressure in front of the turbo-pump running at full speed.  The upper pressure limit can be extended if 
required, with simple modifications to the vacuum system.  Before any ion trapping operations begin, the entire vacuum system is baked for 
two days at 135~$^{\circ}$C in an oven that completely encloses the tank.  After bakeout, the system reaches a base pressure of 
$<3\times10^{-9}$~torr.

\begin{figure}[htb!!]
\begin{center}
\includegraphics[width=13cm]{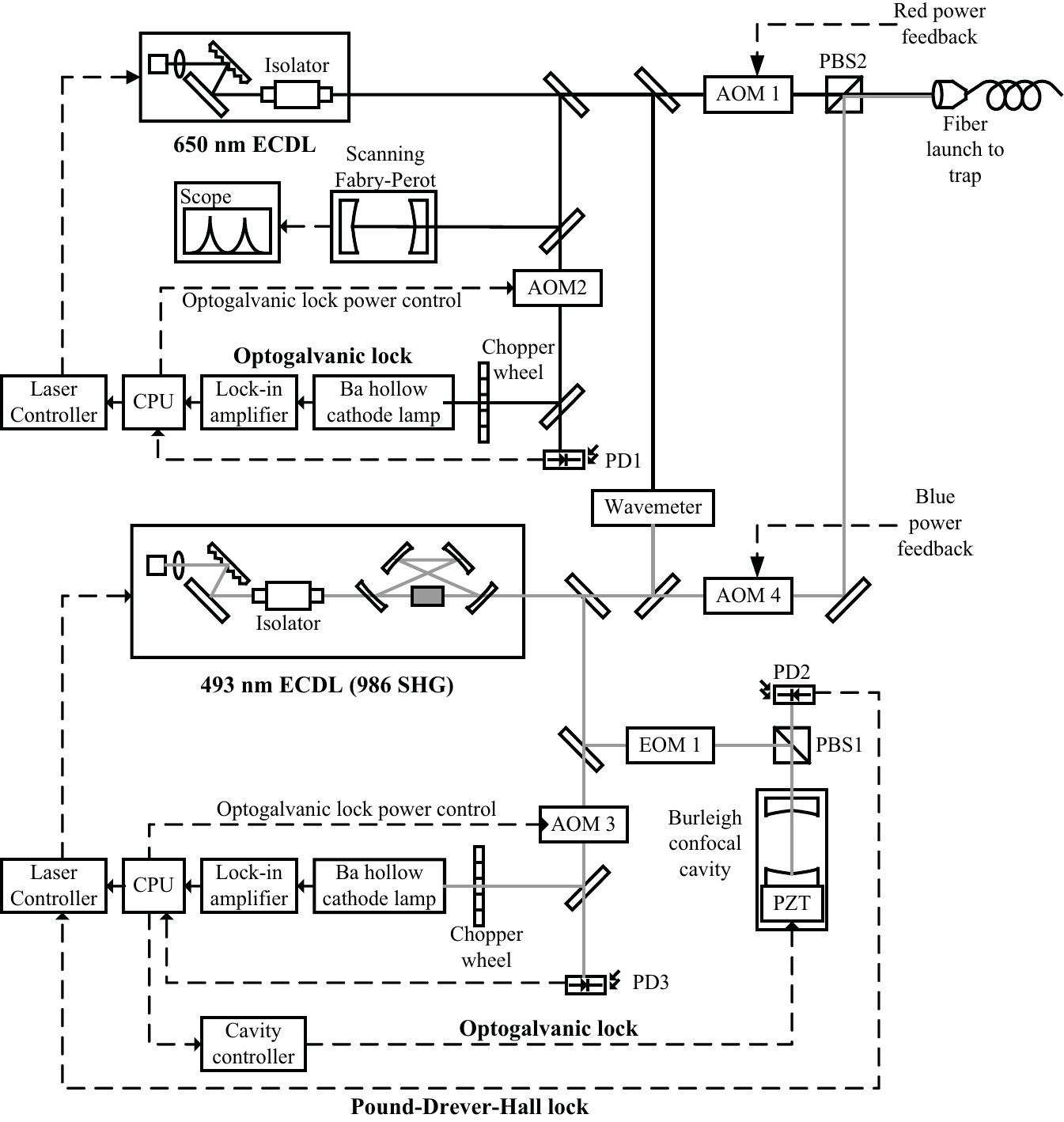}
\vskip 0.1cm
\caption{Schematic view of the laser setup.}
\label{laser_setup}
\end{center}
\end{figure}

Fluorescence from a trapped Ba ion is induced, following the classic ``shelving'' scheme~\cite{Neuhauser82}, by resonant lasers cycling the
ion between the 6S$_{1/2}$ (ground), 6P$_{1/2}$ (excited), and 5D$_{3/2}$ (metastable) states.  The ion undergoes spontaneous emission when 
in the 6P$_{1/2}$ state, emitting either a 493~nm or 650~nm photon.  The 493~nm fluorescence photons are the signal collected for this 
experiment.  The 6S$_{1/2}\leftrightarrow6$P$_{1/2}$ transition at 493~nm is excited by a frequency-doubled external-cavity diode 
laser (ECDL)\footnote{TOPTICA SHG 100}.  The 6P$_{1/2}\leftrightarrow5$D$_{3/2}$ transition at 650~nm is excited by a ECDL\footnote{TOPTICA 
DL 100}.  
20~mW (10~mW) of 493~nm (650~nm) light is available for spectroscopy, far in excess of that required to observe a single ion.  The blue 
laser is frequency stabilized to $\sim 20$~MHz (relative) using an Invar Fabry-Perot reference cavity. Long term absolute frequency 
stabilization is achieved by locking both lasers to a hollow-cathode Ba lamp\footnote{Perkin-Elmer Ba Lumina HCL model N305-0109}.  A 
schematic of the laser setup is shown in fig. \ref{laser_setup}. 

The laser systems reside on a vibration isolated, optics table in a dust controlled environment, completely separated from the linear ion trap vacuum system. 
Both lasers are fed into one single-mode fiber, which is routed over an arbitrary distance to the ion trap system.  The output beams are 
coupled into the ion trap via injection optics, consisting of an aspheric focusing lens, an iris to reduce beam halo, and a beam-steering 
mirror.
The beams are directed into the vacuum system through an anti-reflection (AR) coated window\footnote{``Super-V" AR coating (99.98
\% transmission at 493~nm) by OptoSigma} along the longitudinal axis of the trap, from the end of the trap closest to S$_{14}$.  An 
additional aperture inside the vacuum system aids in beam alignment and further halo reduction.  Due to the chromaticity of the aspheric 
lens, the beam waists are separated by 260~mm, which is on the order of the beams Rayleigh length.  The 493~nm and 650~nm waists at the trapping segment ($S_{14}$) are 370~$\mu$m and 570~$\mu$
m, respectively.  The laser powers at the injection region are sampled in real-time by photodiodes.  These powers are fed-back to acousto-optic modulators before the fiber input on the laser 
table, and used to keep the injected beam powers stable to $\sim$~1~\% indefinitely. This configuration is found to suppress background 
laser light levels to the level required for observing a single ion.

The fluorescence from a trapped ion is detected by an electron-multiplied CCD camera (EMCCD), sensitive to single photons\footnote{Andor iXon
$^{\rm EM}$+}.    The fluorescence is imaged onto the EMCCD by a 64~mm working distance microscope\footnote{Infinity InFocus KC with IF4 
objective}, with a numerical aperture of 0.195.  The outer lens of the microscope objective is placed close to S$_{14}$, outside a re-entrant 
vacuum window (see fig. \ref{vac_overview}).  A spherical mirror inside the vacuum tank, directly behind S$_{14}$, reflects fluorescence 
light back through the trap that would otherwise be lost.  This roughly doubles the fluorescence light collection efficiency.  A filter in front of 
the EMCCD attenuates the 650~nm light by more than 99~\% while transmitting $\sim~85\%$ of the 493~nm light.   The light collection 
efficiency of the system is estimated to be $10^{-2}$, including the 90~\% quantum efficiency of the EMCCD.

\section{Trap operation and results}

After the initial pump-down and bakeout, the trap is loaded with ions by ionizing neutral Ba in the central region of S$_3$ (see fig.~\ref{linear_trap}). Barium is chemically produced, after the system has been pumped to 
good vacuum, by heating a ``barium dispenser''\footnote{SAES: http://www.saesgetters.com/default.aspx?idpage=460} loaded with BaAl$_4$-Ni,
and depositing Ba on a Ta foil. The foil can be resistively heated repeatedly, producing a Ba vapor in S$_3$, which is ionized by a 500~eV electron beam from an electron gun\footnote{Kimball Physics FRA-2X1-2016}.  A 32-gauge thermocouple on the Ta foil is used to control the temperature of the foil, regulating the amount of Ba emitted. Before loading ions into the trap, a buffer gas is introduced into the system.  To load small numbers of ions ($<$~10), the oven is operated at $100~^{\circ}$C and the e-gun pulsed for 10~s. Ions are cooled by the buffer gas, and trapped at $S_{14}$ as discussed earlier.  The trap can also be loaded by turning on one of the cold cathode vacuum gauges.  This effect is explained by the possible emission of electrons and ions from the gauge, which is presumably coated with Ba from the initial Ba source creation process. This should not be considered a background for Ba tagging in EXO, as the trap used in EXO will not be heavily contaminated with Ba, nor will vacuum gauges be operated during the tagging process.

\begin{figure}[htb!!]
\begin{center}
\includegraphics[width=12cm]{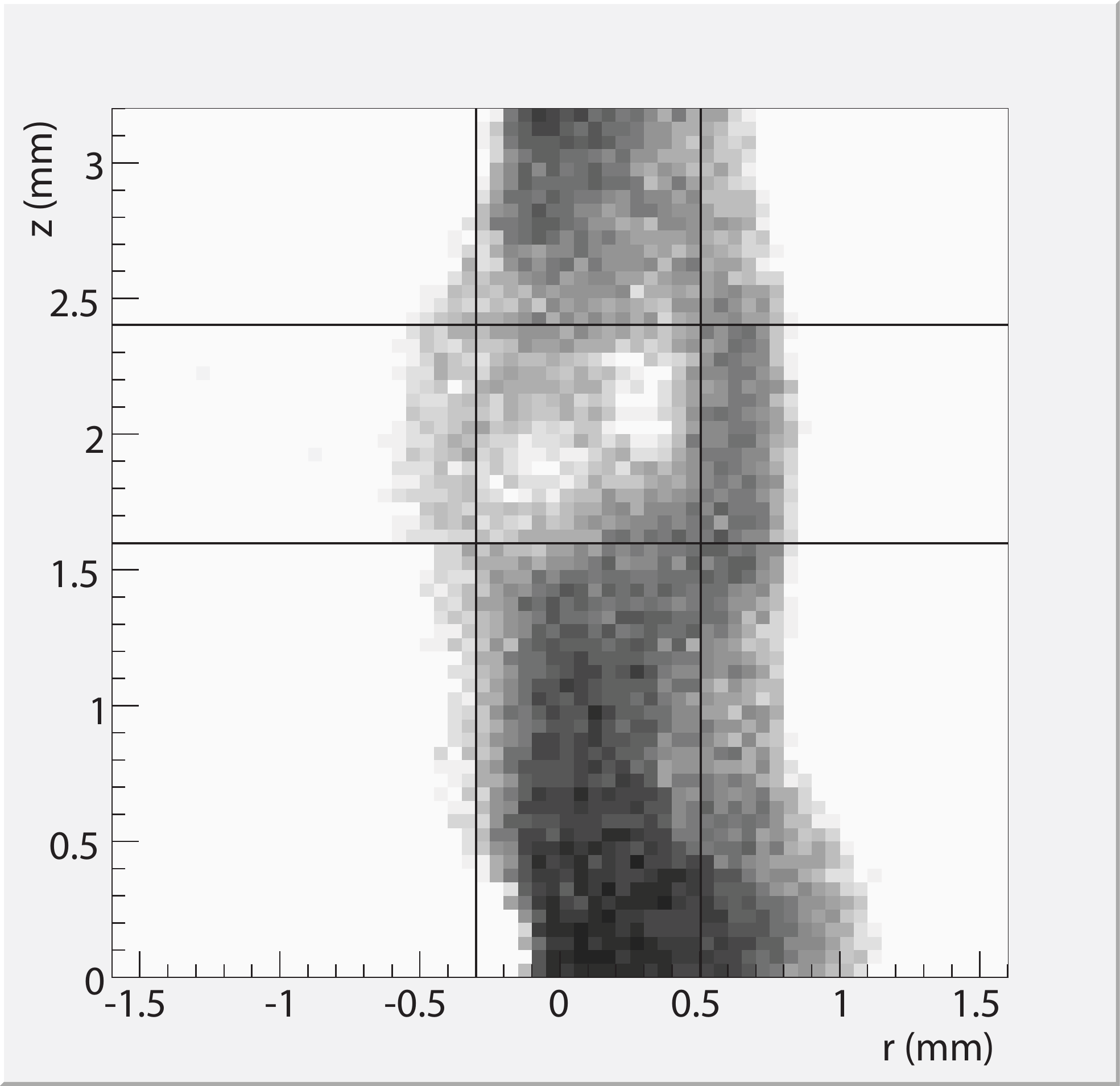}
\vskip 0.1cm
\caption{Greyscale picture of three ions contained in segment 14 of the trap. The image was taken at 
$5\cdot10^{-4}$~torr He. The signal in the region of interest (black box) was integrated over 20 seconds with the EMCCD. }
\label{ion_cloud}
\end{center}
\end{figure}

Fig.~\ref{ion_cloud} shows a grayscale image of the fluorescence from three ions in the trap, imaged by the EMCCD over 20~s.  The DC 
potentials are set as shown in fig.~\ref{linear_trap}.  The brightness of each pixel is proportional to the number of collected photons.  
The cloud of white pixels (encompassed by the black square) is fluorescence from three ions trapped in S$_{14}$. The cloud has two lobes, 
due to a slight misalignment of the spherical mirror behind S$_{14}$.   The white vertical bands on either side of the cloud are due to 
laser light scattering off the electrodes, which is at the same wavelength as the fluorescence photons.  Precise alignment of the laser 
beams is required to minimize scattered laser light and optimize the signal to noise in the region of interest, in order to observe the 
fluorescence from a single ion in the trap.

\begin{figure}[htb!!]
\begin{center}
\includegraphics[width=12cm]{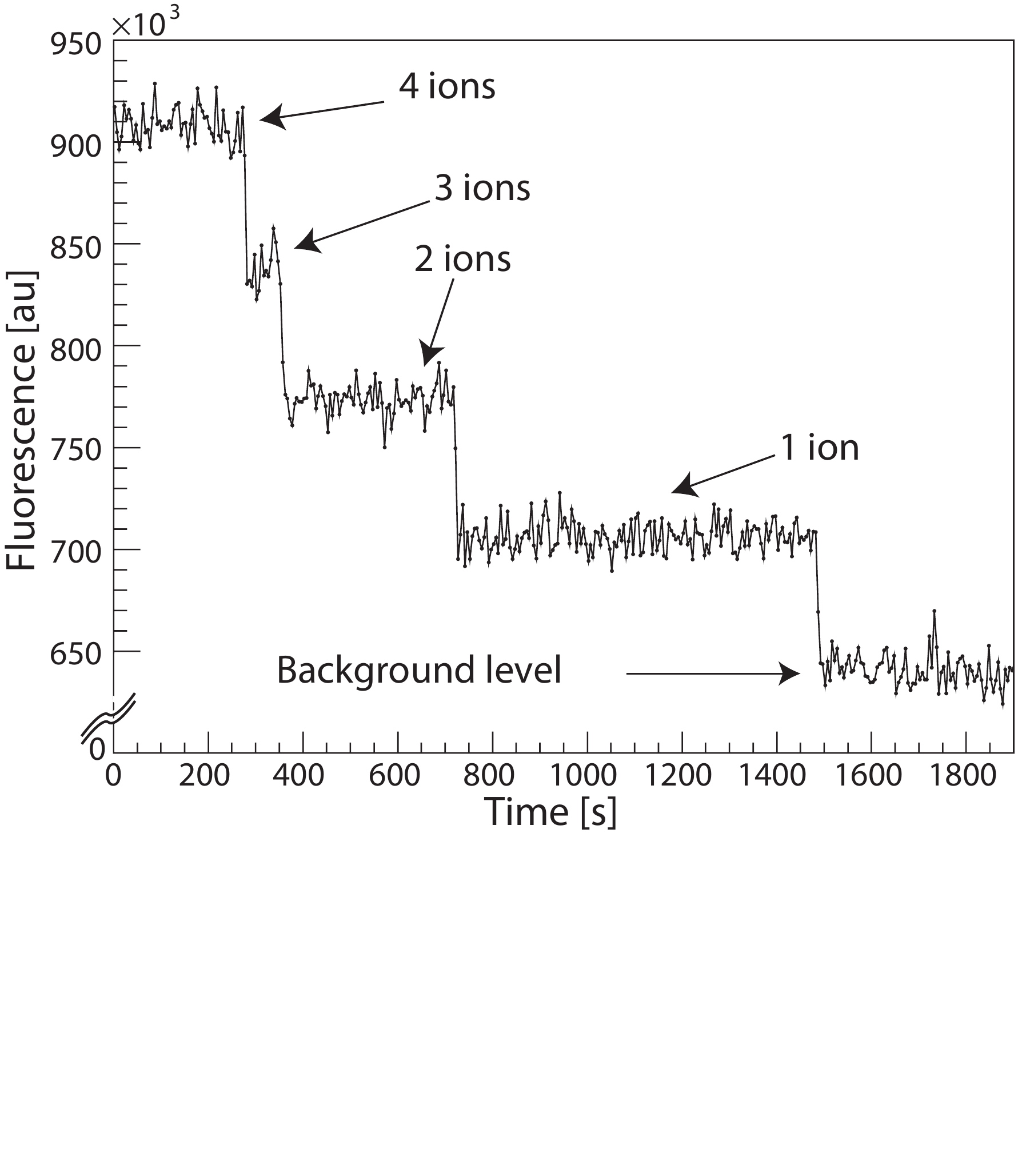}
\vskip 0.1cm
\caption{Time series of the 493 nm ion fluorescence rate in the trap
at 4.4$\cdot10^{-3}$ torr He. Ions unload, causing clear quantized drops in the fluorescence rate.
Each point represents 5 s of integration with the EMCCD \cite{prl}.}
\label{time_series}
\end{center}
\end{figure}

A time-series of the Ba ion fluorescence signal is shown in fig.~\ref{time_series}~\cite{prl}, starting with four ions loaded into the trap at $4.4\times10^{-3}$~torr He.  The x-axis is in seconds, and the y-axis pepresents the fluorescence in arbitrary units of EMCCD counts.  Each data point in this series is the sum of the pixels the the square box in fig.~\ref{ion_cloud}, after 5~s of integration by the EMCCD.  The y-axis is zero-suppressed due to the large EMCCD pedestal.  Over time, ions spontaneously eject from the trap due to RF heating, and possibly ion-ion Coulomb interactions.  As individual ions eject, the fluorescence signal decreases in quantized steps. 
The difference in the fluorescence signals for the single steps is constant within 4\%, clearly establishing the capability of the system in detecting single ions.

A high signal-to-noise ratio of the fluorescence from a single trapped ion will be required to confirm a $0\nu\beta\beta$ decay.  The signal-to-noise ratio for this purpose is defined as 
\begin{center}
\begin{equation}
S/N=\frac{\left<R_I\right> - \left<R_B\right>}
         {\sqrt{\frac{\sigma_I^2\Delta t}{t_I}+\frac{\sigma_B^2\Delta t}{t_B}}}
\label{eqn:SN}
\end{equation}
\end{center}
where $\left<R_I\right>$ and $\left<R_S\right>$ are the average single-ion fluorescence and background rates, $\sigma_I$ and $\sigma_B$ are the Gaussian widths of the single-ion fluorescence and background rates, $t_I$ and $t_B$ are the total single-ion fluorescence and background rate observation times, and $\Delta t$ is the integration time of single measurement (hence $t_I / \Delta t$ and $t_B / \Delta t$ are the number of measurements for the signal and the background, respectively).  This metric assumes that both the single ion fluorescence and background rates follow Gaussian statistics.  If the two integration times are equal ($t_I=t_B=t$), eqn.~\ref{eqn:SN} becomes
\begin{center}
\begin{equation}
S/N = \frac{\left<R_I\right>-\left<R_B\right>}{\sqrt{\sigma_I^2+\sigma_B^2}}\sqrt{\frac{t}{\Delta t}}= k\sqrt{t}
\label{eqn:SN2}
\end{equation}
\end{center}
where the signal and background rates have been absorbed into the constant $k$, in units of Hz$^{-1/2}$.  The signal-to-noise ratio of the fluorescence from an individual ion increases with the square root of the measurement time, or equivalently number of measurements.  For the single ion fluorescence and background rates in fig.~\ref{time_series}, $k=2.75$~Hz$^{-1/2}$, so that $S/N=18$ for a 60s measurement time.  Similar values are found in the cases of Ar, and He/Xe mixtures as buffer gases.
Drifts in the laser beam position at S$_{14}$ may lead to fluctuations in the background rate, $R_B$, that are not accounted for by the assumptions of Gaussian statistics made here.    Such drifts are caused primarily by temperature variations in the lab, affecting the alignment  of the trap injection optics.    In the setup used for the data presented here, no provision is made for the temperature stabilization of such optics that drift by as much as $\pm 2^{\circ}$C on a daily cycle.   Even under these non-ideal conditions, the system is stable enough to apply the $S/N$ description of eqn.~\ref{eqn:SN} for periods of minutes, much longer than required for non-ambiguous single Ba ion identification.    Temperature stabilization of the injection optics capable of $\pm 0.1^{\circ}$C would be straightforward to implement and would reduce non-Gaussian background fluctuations to a negligible level. 

The lifetimes of single Ba ions in the trap at different He pressures have been measured and reported elsewhere~\cite{prl}.    In the same paper, the destabilizing effects of Xe contaminations are also studied and modeled.    It is found that a sufficient partial pressure of He in the trap can counter the effects of Xe, and provide lifetimes that are sufficient for single-ion detection with very high significance.

\section{Summary}

A linear RFQ ion trap, designed and built within the R\&D program towards the EXO experiment, is described.  The trap is capable of confining individual Ba ions for observation by laser spectroscopy, in the presence of light buffer gases and low Xe concentrations.  Single trapped Ba ions are observed, with a high signal-to-noise ratio.  A similar trap will be used to identify single Ba ions produced in the $0\nu\beta\beta$ decay of $^{136}$Xe in the EXO experiment.  The successful operation of this trap, as described here, is one of the cornerstones of a full Ba tagging system for EXO, which will lead to a new method of background suppression in low-background experiments.  In parallel, several systems to capture single Ba ions in LXe, and transfer them into the ion trap are under development.

{\bf Acknowledgements}\\
\\
This work was supported, in part, by DoE grant FG03-90ER40569-A019 and by private funding from Stanford University. We also gratefully acknowledge a substantial equipment donation from the IBM corporation, as well as very informative discussions with Guy Savard.



\newpage


\begin{thebibliography}{00}




\bibitem{oscillations} Y.~Fukuda, T.~Hayakawa, E.~Ichihara, K.~Inoue,{\it et al.}, Phys. Rev. Lett. 81 1562 (1998).\\
 M.H.~Ahn, E.~Aliu, S.~Andringa, S.~Aoki {\it et al.}, Phys. Rev. D 74, 072003 (2006).\\
 Q.R.~Ahmad, R.C.~Allen, T.C.~Andersen, J.D~Anglin {\it et al.}, Phys. Rev. Lett 89 011301 (2002).\\
 T.~Araki, K.~Eguchi, S.~Enomoto, K.~Furuno, Phys. Rev. Lett. 94 081801 (2005).\\
 B.T.~Cleveland,T.~Daily, R.~Davis, Jr., J.R.~Distel {\it et al.}, Astrophys. J. 496, 505 (1998).\\
 J.N.~Abdurashitov, V.N.~Gavrin, S.V.~Girin, V.V.~Gorbachev {\it et al.}, Phys. Rev. C 60, 055801 (1999).\\
 W.~Hampel, J.~Handt, G.~Heusser, J.~Kiko {\it et al.}, Phys. Lett. B 447, 127 (1999).\\
 D.G.~Michael, P.~Adamson, T.~Alexopoulos, W.W.M.~Allison {\it et al.}, Phys. Rev. Lett. 97, 191801 (2006).\\

\bibitem{betadecay} A.~Osipowicz, H.~Blumer, G.~Drexlin, K.~Eitel {\it et al.}, arxiv hep-ex/0109033.\\ 
 A.~Minfardini, C.~Arnaboldi, C.~Brofferio, S.~Capelli {\it et al.}, Nucl. Instr. Meth. A 559 346 (2006).

\bibitem{Elliott-Vogel} S.~Elliott, P.~Vogel, Ann. Rev. Nucl. Part. Sci. 52, 11551 (2002).

\bibitem{Majorana} E.~Majorana, Nuovo Cim. 14 (1937) 171.

\bibitem{EXO_whitepaper} M.~Danilov, R.~DeVoe, A.~Dolgolenko, G.~Giannini {\it et al.}, Phys. Lett. B 480 (2000) 12.\\
M.~Breidenbach, M~Danilov, J.~Detwiler {\it et al.}, R\&D proposal, Feb 2000, Unpublished.

\bibitem{Denison} D.~Denison, J. Vac. Sci. Tech. 8, 266 (1971). 

\bibitem{paul} W.~Paul, Rev. Mod. Phys. 62, 531 (1990).

\bibitem{March}R. Marchetal., {\it Quadrupole Ion Trap Mass Spectrometry}, Wiley-Interscience, (2005).

\bibitem{Mathieu} N.~McLachlan, Theory and Application of Mathieu Functions (Dover, 1964).

\bibitem{prl} M.~Green, J.~Wodin, R.~deVoe, P.~Fierlinger {\it et al.}, arXiv:physics/0702122 (2007), submitted to PRL.

\bibitem{Waldman_Wodin} S.~Waldman, PhD Thesis, Stanford University 2005.\\
                        J.~Wodin, PhD Thesis, Stanford University 2007.

\bibitem{kim} K.~Taeman, PhD Thesis, McGill University 1997.

\bibitem{Neuhauser82} W.~Neuhauser, M.~Hohenstatt, P.~Toscheck, H.~Dehmelt, Phys. Rev. Lett. 41 (1978) 233.
\end{thebibliography}
\end{document}